\documentclass[aps,prb,amsmath,amssymb,twocolumn,10pt]{revtex4-1}

  \usepackage{times,mathptmx}
  \usepackage{graphicx}
  \usepackage{bm}
  \usepackage{color}

\begin{document}
\title{Effect of Coulomb Interactions on the Electronic and Magnetic Properties of Two-Dimensional CrSiTe$_3$ and CrGeTe$_3$ Materials}

\author{Sungmo Kang}
\author{Seungjin Kang}
\author{Jaejun Yu}

\email{jyu@snu.ac.kr}
\affiliation{Center for Theoretical Physics, Department of Physics and Astronomy, Seoul National University, Seoul 08826, Korea}

\date{\today}

\begin{abstract}
  We investigate the electronic and magnetic structures of two-dimensional transition metal trichalcogenide CrSiTe$_{3}$ and CrGeTe$_{3}$ materials by carrying out first-principles calculations. The single-layer CrSiTe$_3$ and CrGeTe$_3$ are found to be a ferromagnetic insulator, where the presence of the strong $dp\sigma$-hybridization of Cr $e_{\mathrm{g}}$-Te $p$ plays a crucial role for the ferromagnetic coupling between Cr ions. We observe that the bandgaps and the interlayer magnetic order vary notably depending on the magnitude of on-site Coulomb interaction $U$ for Cr $d$ electrons. The bandgaps are formed between the Cr $e_{\mathrm{g}}$ conduction bands and the Te $p$ valence bands for both CrSiTe$_3$ and CrGeTe$_3$ in the majority-spin channel. The dominant Te $p$ antibonding character in the valence bands just below the Fermi level is related to the decrease of the bandgap for the increase of $U$. We elucidate the energy band diagram, which may serve to understand the electronic and magnetic properties of the $ABX_3$-type transition metal trichalcogenides in general.

\end{abstract}


\maketitle

\section{Introduction}
\label{sec:introd}

Transition metal dichalcogenides (TMDC) in their atomically thin two-dimensional (2D) forms exhibit a wide range of electronic, optical, mechanical, chemical and thermal properties. In particular, their tunable bandgap properties depending on the number of layers make this class of materials as a candidate for future electronics and optoelectronics applications.\cite{Wang:2012aa} Due to the presence of transition metal atoms, however, the emergence of magnetism in 2D crystals has opened up interesting possibilities. For example, chromium triiodide (CrI$_3$) was suggested as an ideal candidate for 2D magnets exfoliated from easily cleavable single crystals of CrI$_3$, which is a layered and insulating ferromagnet with a Curie temperature of 61 K.\cite{McGuire:2015aa} A recent observation of ferromagnetism has demonstrated its layer dependence down to the monolayer limit.\cite{Huang:2017aa} Along with TMDC, another class of layered transition-metal trichalcogenides (TMTC) with the chemical formula $ABX_3$ ($A$ = Mn, Cr; $B$ = Si, Ge; $X$ = S, Se, Te) have attracted interest as potential candidates for two-dimensional magnets.\cite{sivadas2015magnetic}

Although these $ABX_3$-class of TMTC materials have been studied for many decades,\cite{Wiedenmann:1981aa,Brec:1986aa,Siberchicot:1996aa,Wildes:1998aa,Joy:1992aa,Takano:2004aa} their electronic and magnetic structures as well as mechanism for magnetic ordering are not clearly understood yet. For instance, CrSiTe$_{3}$, one of the TMTC materials, is well known as a candidate for a 2D ferromagnetic (FM) semiconductor. The Curie temperatures were reported to increase as the number of layers is reduced.\cite{lin2016ultrathin,williams2015magnetic} On the other hand, however, there exist conflicts on the predicted magnetic ground states. Different magnetic ground states are proposed for bulk and single-layer CrSiTe$_{3}$.\cite{chen2015strain,casto2015strong} Further, a magnetic phase transition was suggested to occur under the tensile strain.\cite{sivadas2015magnetic,chen2015strain}

As a step toward understanding the origin of ferromagnetism in the $ABX_3$-class materials, we investigate the electronic and magnetic structures of 2D TMTC CrSiTe$_{3}$ and CrGeTe$_{3}$ materials by carrying out first-principles calculations. We performed total energy calculations for various magnetic configurations in single-, bi-, and triple-layers as well as bulk CrSiTe$_{3}$ and CrGeTe$_{3}$ including their full structural optimizations. We also examine the effect of on-site Coulomb interactions $U$ for Cr $d$ electrons by monitoring the bandgap and magnetic order. The results show an unusual behavior of bandgap as well as magnetic order depending on $U$, which may provide a clue to the understanding of the electronic and magnetic properties of the $ABX_3$-type TMTC materials in general.

\section{Methods}
\label{sec:methods}
The first-principles calculations were performed by using the density functional theory (DFT) within the generalized gradient approximation GGA+$U$ method. To obtain band structures and projected density of states, we use the OpenMX code\cite{openmx,Ozaki:2005aa} which employ localized orbital bases, especially with the GGA exchange-correlation functional in the parameterization of Perdew, Burke, and Enzerhof (PBE)\cite{perdew1996generalized}. We use the effective on-site Coulomb interaction $U_\mathrm{eff}=U-J$ in a Dudarev implementation\cite{Dudarev:1998aa,Han:2006aa} to treat the localized Cr $d$ states throughout the calculations. We obtain the electronic and magnetic structures by varying the $U_{\mathrm{eff}}$ values from 0.0 to 3.0 eV, which will be called as $U$, for simplicity, from now on. {To examine the $U$-dependence of bandgap, we also carried out the hybrid functional calculations as a reference by using the HSE06 exchange-correlation functional\cite{Krukau:2006aa} as implemented in the VASP package.\cite{Kresse:1996aa}}

To simulate a single or few layers of 2D Cr$B$Te$_3$ ($B$ = Si, Ge) systems, we make use of a slab geometry with 20 {\AA} vacuum in-between the slab layers, where each layer consists of a 2D honeycomb lattice with the Cr$_2$B$_2$Te$_6$ unit cell. The cutoff energy of 500 Ry is used for the real and momentum space grids and the \textbf{k}-mesh of $10\times 10\times 1$ for the Brillouin zone integration. The lattice structures are relaxed under the constraint of $C_3$ rotation symmetry until the residual forces converge within $10^{-4}$ in the atomic unit.

\begin{figure}
	\includegraphics[width=0.9\linewidth]{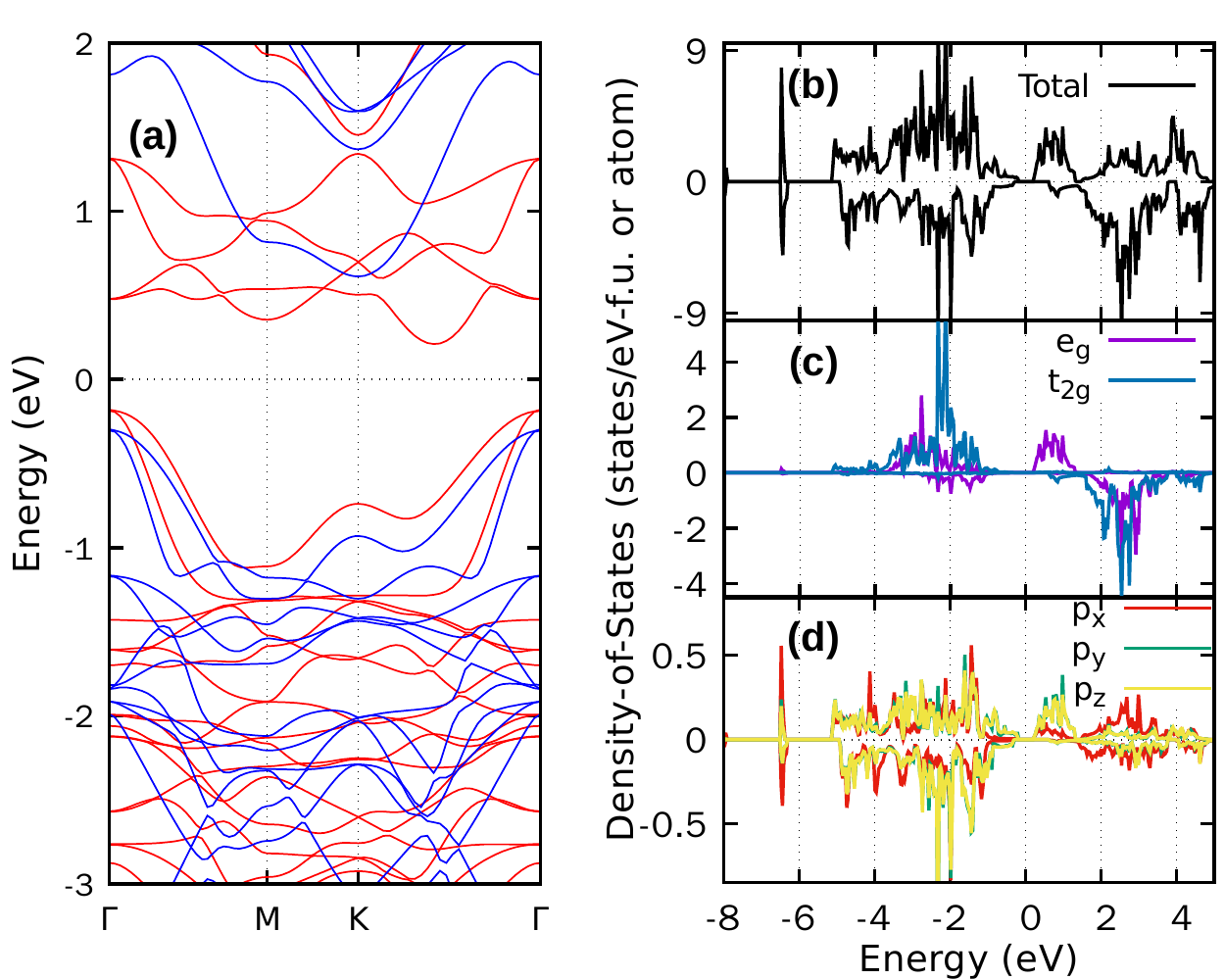}
	\caption{(Color online) (a) Band structure, (b) total, (c) Cr- and (d) Te-projected density-of-state (pDOS) of ferromagnetic single-layer CrSiTe$_3$ at $U=1.5$ eV. In the band structure plot, the majority-spin (minority-spin) bands are marked by the red (blue) lines, respectively. In the pDOS plots, the upper panel represents for the majority-spin (spin-up) components of the pDOS and the lower panel for the minority (spin-down) components, where the Fermi level ($E_{\mathrm{F}}$) is set to zero.}
	\label{fig:band-pdos-CST}
\end{figure}

\section{Results and Discussion}
\label{sec:results}

\subsection{Electronic structure and magnetic properties of single-layer CrSiTe$_{3}$}
\label{sub:electronic}

We carried out first-principles calculations for the ground states of Cr$B$Te$_{3}$ ($B$ = Si, Ge). Our results of the electronic band structures and the magnetic ground states for the single-layer and bulk systems are in general agreement with the previous works.\cite{li2014crxte,chen2015strain,casto2015strong,sivadas2015magnetic,lin2016ultrathin} To calculate the electronic band structures and projected density of states (pDOS), we adopt the on-site Coulomb interaction parameter of $U=1.5$ eV. More discussion on the choice of $U$ will be made in Section~\ref{sub:magnetic-ground}.

Both single-layer CrSiTe$_3$ and CrGeTe$_3$ are determined to be a FM insulator. Both bulk CrSiTe$_3$ and CrGeTe$_3$ have the same space group of R3 (No.148) in common with other $ABX_3$ TMTC. The optimized in-plane lattice constants are $a=6.78$ {\AA} and 6.86 {\AA} for the single-layer CrSiTe$_3$ and CrGeTe$_3$, respectively. For the bulk structures with the $AB$ stacking sequence, the lattice parameters are determined to be $a=6.8$ {\AA} and $c=13.4$ {\AA} for CrSiTe3$_3$ and $a=6.9$ {\AA} and $c=13.2$ {\AA} for CrGeTe$_3$.

Figure~\ref{fig:band-pdos-CST} shows the spin-polarized band structure and pDOS for the FM single-layer CrSiTe$_3$ with $U=1.5$ eV. Since CrGeTe$_3$ exhibits similar features of the valence and conduction bands except for the states related to Ge, here we focus on the electronic structure of CrSiTe$_3$ only. The prominant features of the CrSiTe$_3$ electronic structure are the \emph{empty} $dp\sigma$-hybridized antibonding bands of Cr $e_{\mathrm{g}}$-Te $p$ at $\sim 1$ eV above the Fermi level ($E_{\mathrm{F}}$) and the \emph{fully occupied} Cr $t_{\mathrm{2g}}^{3\uparrow}$ bands at about $-2$ eV below $E_{\mathrm{F}}$. The unoccupied spin-down (minority-spin) bands of Cr $t_{\mathrm{2g}}$ are located at about 2 eV above $E_{\mathrm{F}}$, indicating a large exchange splitting between the localized Cr $t_{\mathrm{2g}}$ orbitals. Thus, the local magnetic moment of each Cr atom is 3.87 $\mu_{\mathrm{B}}$, where the extra contribution of 0.87 $\mu_{\mathrm{B}}$ comes from the $dp\sigma$ bonding states of Cr $e_{\mathrm{g}}$-Te $p$. In fact, this Cr-Te $dp$-hybridization gives rise to the Te $p$ holes with an opposite spin polarization of $-0.3$ $\mu_{\mathrm{B}}$ per Te atom so that the \emph{total} FM moment per CrSiTe$_3$ unit-cell remains 3 $\mu_{\mathrm{B}}$. In addition, the single-ion anisotropy energy is found to be about 0.77 meV and 0.31 meV per Cr atom with an easy axis perpendicular to the layer for the single-layer CrSiTe$_3$ and CrGeTe$_3$, respectively. It indicates that both CrSiTe$_3$ and CrGeTe$_3$ are Ising-like ferromagnets, in agreement with previous experiment\cite{sivadas2015magnetic,casto2015strong} and calculation\cite{Zhuang:2015aa} results.

The presence of the hybridized Te $p$ holes generated by the strong $dp\sigma$-hybridization between Cr $d$ and Te $p$ orbitals plays a crucial role in the FM-coupling mechanism in this class of TMTC materials. Apart from the regular superexchange contributions, which may be valid for the fully occupied Cr $t_{\mathrm{2g}}^{3\uparrow}$ states, the itinerant holes residing in the Te $p$ ligands are coupled to their neighboring Cr spins antiferromagnetically, mediating the FM ordering of Cr local moments. This mechanism shares a common feature with the Zener's mechanism\cite{Zener:1951aa} where an effective exchange interaction is generated by the $sd$-hybridization instead of the $pd$-hybridization. Therefore, to stabilize the FM ordering of Cr spins, it is essential to have the energy gain by the negative polarization of the Te $p$ state, which is considered as a relaxation of the non-magnetic elements.\cite{Kanamori:2001aa}

\begin{figure}
	\includegraphics[width=0.9\linewidth]{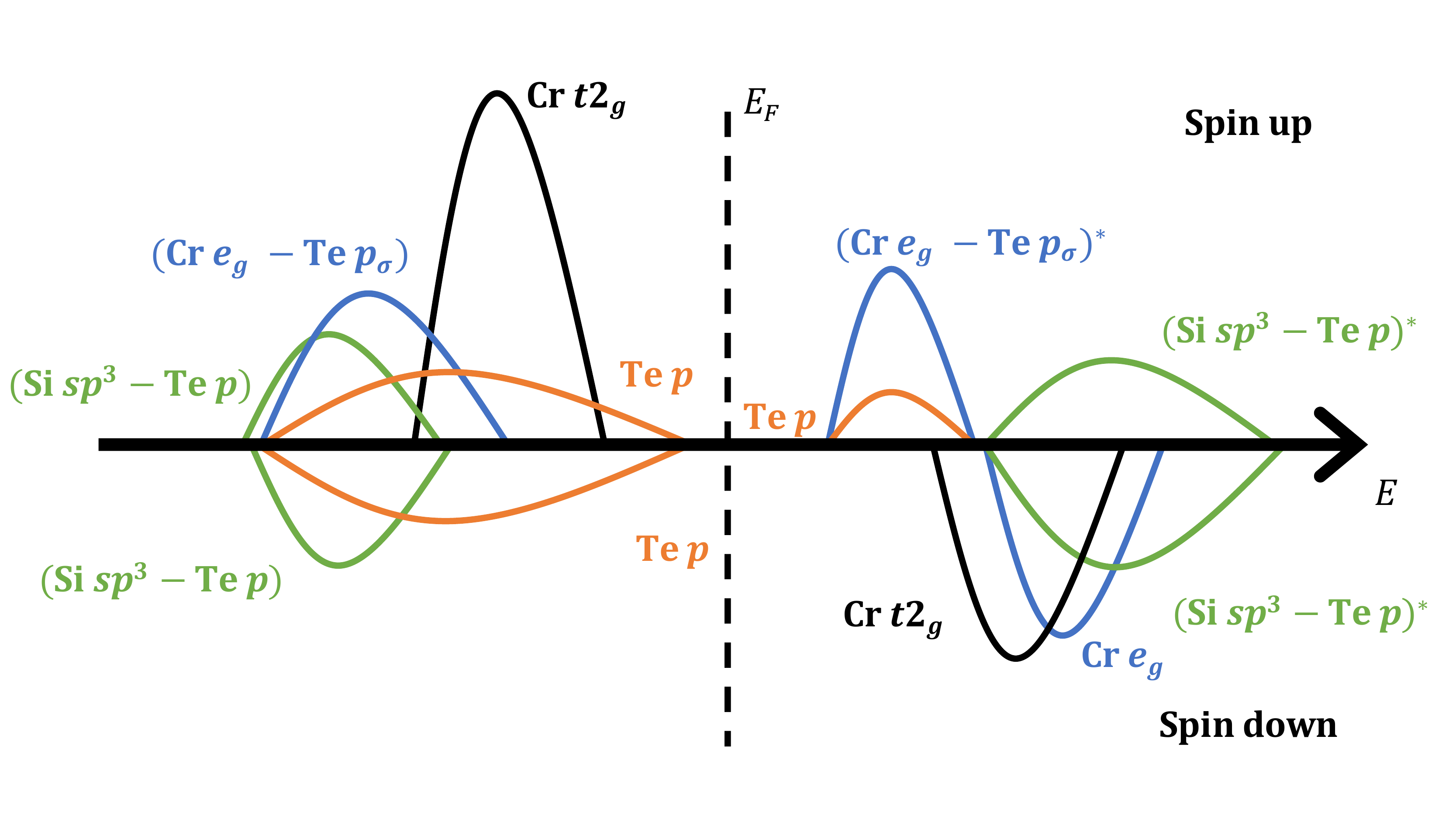}
  \caption{(Color online) Schematic energy diagram of the CrSiTe$_{3}$ electronic structure. The primary features are the bonding and antibonding bands of (Cr $e_{\mathrm{g}}$-Te $p\sigma$) and (Si $sp^3$-Te $p$) hybridized states, whereas the localized Cr $t_{\mathrm{2g}}$ bands are split into the spin-up and spin-down channels across the Fermi level ($E_{\mathrm{F}}$).}
	\label{fig:energy-diagram}
\end{figure}

To help the understanding of the electronic structure of TMTC, we present a schematic energy diagram for CrSiTe$_3$ in Fig.~\ref{fig:energy-diagram}. This diagram may serve as a representative picture for the electronic configuration of 2D TMTC materials. As we discussed above, each Cr $e_{\mathrm{g}}$ orbital form a bonding and antibonding pair of (Cr $e_{\mathrm{g}}$-Te $p_{\sigma}$) states, whereas the weak $dp\pi$ hybridization leads to the localizaed Cr $t_{\mathrm{2g}}$ states. One notable feature is that the Si 3$s$ level is located at $-6.5$ eV below $E_{\mathrm{F}}$, which is not shown in Fig.~\ref{fig:band-pdos-CST}. Since the Si atom has the tetrahedral coordination surrounded by another Si atom and 3 Te atoms, the Si $sp^3$ hybrid orbitals can make a strong bonding and antibonding pair of (Si $sp^3$-Te p). Hence, the bandgap in the spin-up channel is formed between the Cr $e_{\mathrm{g}}$ conduction band and the Te $p$ valence band for both CrSiTe$_3$ and CrGeTe$_3$. The principal components near the top of the valence bands consist of the anti-bonding Te $p$-Te $p$ character, while the conduction bands are from the anti-bonding Cr $e_{\mathrm{g}}$-Te $p\sigma$ orbitals.

\subsection{On-site $U$ and magnetic ground states of Cr$B$Te$_{3}$ ($B$ = Si, Ge)}
\label{sub:magnetic-ground}

\begin{figure}
	\includegraphics[width=0.9\linewidth]{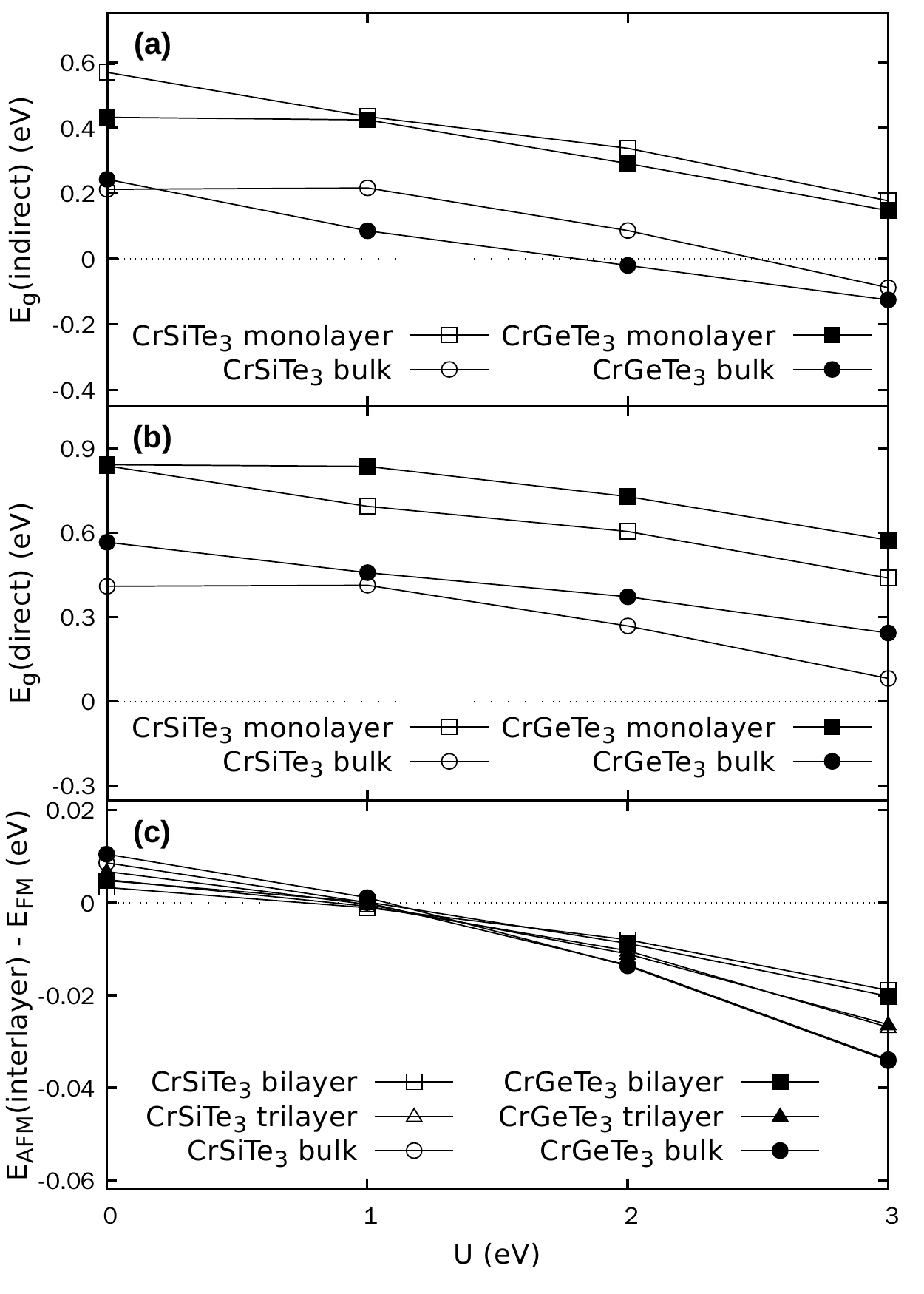}
	\caption{(a) Indirect and (b) direct bandgaps for the monolayer and bulk CrSiTe$_3$ and CrGeTe$_3$ systems and (c) total energy differences between the A-type (i.e., interlayer-antiferro) antiferromagnetic (AFM) and the ferromagnetic (FM) for bilayer, trilayer and bulk CrSiTe$_3$ and CrGeTe$_3$ depending on the on-site Coulomb interaction parameter $U$.}
	\label{fig:u-dependence}
\end{figure}

From the results of calculations with varying $U$, we observe an interesting but still critical behavior of bandgap as well as magnetic order depending on the on-site Coulomb interactions for Cr $d$ orbital states. As illustrated in Fig.~\ref{fig:u-dependence}(a) and (b), the change of the indirect and direct bandgaps with varying on-site $U$ parameters demonstrates that the bandgaps are sensitive to the choice of the $U$ values for CrSiTe$_3$ and CrGeTe$_3$ monolayer as well as bulk systems. For example, the $U=0$ calculations show an insulating ground state with finite gaps, while $U=3$ eV predicts a semi-metallic ground state with \emph{negative} indirect gaps for both CrSiTe$_3$ and CrGeTe$_3$. This $U$-dependence can be understood from the electronic structures near $E_{\mathrm{F}}$.

The increase of $U$ pushes down the localized spin-up Cr $t_{\mathrm{2g}}$ level relative to the unoccupied Cr $e_{\mathrm{g}}$-Te $p$ hybridized state. But, the top of the valence bands, consisting mostly of the Te $p$ component, is not affected by the change of $U$. The downward shift of the $t_{\mathrm{2g}}$ level in turn raises the anti-bonding Te $p$ bands. Thus, the increase of $U$ contributes to the relative upward shift of the anti-bonding Te $p$ bands, thereby leading to the decrease of the indirect and direct bandgaps. The smaller bandgaps for the bulk systems is attributed to the large bandwidth of the Te $p$ bands, which reflect the overlap of Te $p$ states across the layers.

Thus, the choice of $U$ for the Cr 3$d$ orbitals is crucial in the determination of their ground state. Along with the change of bandgaps, the $U$-parameters also affect the magnetic ordering between the layers. While the single-layer CrSiTe$_3$ and CrGeTe$_3$ favor the FM ground state, the interlayer magnetic couplings are prone to the on-site Coulomb interaction at the Cr site. Figure~\ref{fig:u-dependence}(c) shows that the FM ground state is stable only for $U<1.0$ eV and the AFM order takes over for $U>1.0$ eV. In the case of 3$d$ transition-metal oxides, $U=3.5$ eV was reported for Cr$_2$O$_3$, for instance, from GGA+$U$ calculations in comparison with experiments.\cite{Wang:2006aa} However, if $U=3.5$ eV were adopted for TMTC, both CrSiTe$_3$ and CrGeTe$_3$ would be a semi-metal with the negative bandgap, which contradicts to the semiconducting behavior observed in experiments.\cite{Ji:2013aa,casto2015strong,lin2016ultrathin} Therefore, in the range of $U<1.5$ eV, we conclude that both bulk CrSiTe$_3$ and CrGeTe$_3$ may have the FM or A-type AFM ground state, where the interlayer AFM coupling can be quite small compared to the intralayer FM couplings. In particular, it is noted in Fig.~\ref{fig:u-dependence}(c) that the interlayer coupling becomes almost zero near $U\approx 1.0$ eV.

{As a reference, we obtained the bandgaps from HSE06 hybrid-functional calculations. The HSE06 indirect and direct bandgaps for the single-layer CrSiTe$_3$ are 0.85 eV and 1.18 eV, which are significantly larger than the $U$=0 eV results of 0.57 eV and 0.84 eV, respectively, as illustrated in Fig.~\ref{fig:u-dependence}. Similarly, the HSE06 indirect (0.77 eV) and direct (1.24 eV) bandgaps for the single-layer CrGeTe$_3$ are larger than the $U$=0 eV results of 0.43 eV and 0.84 eV, respectively. Despite the larger bandgaps, the overall features of the electronic structures of the HSE06 hybrid-functional calculations are consistent with the small $U$ results. Further, a recent spectroscopic measurement study also support the reduced value of $U$.\cite{Fujimori:aa,Lin:2017aa} Since the HSE06 bandgaps for transition metal oxides and chalcogenides have a complication in treating localized 3$d$ electrons,\cite{Marsman:2008aa,Li:2013aa} however, it may require further investigations to understand the origin of such reduced $U$ for TMTC.}

\section{Conclusions}
\label{sec:conclusions}

To understand the electronic and magnetic properties of 2D TMDC materials especially of CrSiTe$_{3}$ and CrGeTe$_{3}$, we performed DFT calculations within the GGA+$U$ method. {The single-layer CrSiTe$_3$ and CrGeTe$_3$ are found to be an FM insulator with a small but finite bandgap for $U< 1.5$ eV.} The total magnetic moment per formula unit is 3 $\mu_{\mathrm{B}}$. However, the local magnetic moment of each Cr atom is determined to be 3.87 $\mu_{\mathrm{B}}$ {for $U=1.5$ eV}, where the extra contribution of 0.87 $\mu_{\mathrm{B}}$ comes from the $dp\sigma$ bonding states of Cr $e_{\mathrm{g}}$-Te $p$. It is remarkable that the $-0.3$ $\mu_{\mathrm{B}}$ spin polarization resides at each Te atom as a result of the strong Cr-Te $dp\sigma$ hybridization. This negative polarization of Te $p$ relative to Cr evidences that the strong $dp\sigma$-hybridization of Cr $e_{\mathrm{g}}$-Te $p$ is crucial for the stabilization of ferromagnetic ordering of Cr ions.

In addition to the presence of Te $p$ holes due to the strong Cr-Te $dp$-hybridization, the role of the on-site Coulomb interaction $U$ for Cr $d$ electrons seems to be different from the case of 3$d$ transition metal oxides. The bandgaps for both CrSiTe$_{3}$ and CrGeTe$_{3}$ decrease significantly as $U$ increases. In fact, the bandgaps are formed between the Cr $e_{\mathrm{g}}$ conduction band and the Te $p$ valence band for both CrSiTe$_3$ and CrGeTe$_3$. The dominant Te $p$ antibonding bands in the valence bands just below the Fermi level is related to the decrease of the bandgap for the increase of $U$. {Besides the $U$-dependent bandgaps, the magnetic ground state is also sensitive to $U$. As illustrated in Fig.~\ref{fig:u-dependence}, the interlayer magnetic coupling in both bulk and multilayers of CrSiTe$_3$ and CrGeTe$_3$ can be ferromagnetic ($U\lesssim 1$ eV) or anti-ferromagnetic ($U\gtrsim 1$ eV). Further, near $U\approx 1$ eV, the energy difference between FM and A-type AFM is negligible, and the magnetic response becomes critical. Thus, the magnetic ordering of the TMTC materials may be sensitive to external fields or strains. We hope that our findings serve for the future experimental measurements, which will help our understanding of electronic and magnetic properties of TMTC materials.}

\begin{acknowledgements}
	We gratefully acknowledge A. Fujimori and Kee Hoon Kim for valuable discussions.	This work was supported by the National Research Foundation of Korea (NRF) (no. 2017R1A2B4007100). JY gratefully acknowledges the support and hospitality provided by the Max Planck Institute for the Physics of Complex Systems, where this work was completed during his visit to the institute.
\end{acknowledgements}


\end{document}